\documentclass[aps,prb,twocolumn,superscriptaddress]{revtex4-1}

\usepackage{booktabs}
\usepackage{graphicx}
\usepackage{empheq}
\usepackage{amsmath}
\usepackage{verbatim}
\usepackage{multirow} 
\usepackage{makecell} 
\newcommand{\angstrom}{\textup{\AA}}

\begin{document}


\title{Combined Molecular and Spin Dynamics Simulation of BCC Iron with Vacancy Defects}


\author{Mark Mudrick}
\email[]{mmudrick@uga.edu}
\affiliation {Center for Simulational Physics, The University of Georgia, Athens, Georgia 30602, USA}

\author {Markus Eisenbach}
\affiliation{Oak Ridge National Laboratory, Oak Ridge, Tennessee 37831, USA}

\author {Dilina Perera}
\affiliation {Department of Physics and Astronomy, Texas A\&M University, College Station, Texas 77843, USA}

\author {David P. Landau}
\affiliation {Center for Simulational Physics, The University of Georgia, Athens, Georgia 30602, USA}


\date{\today}

\begin{abstract}
Utilizing an atomistic computational model which handles both translational and spin degrees of freedom, combined molecular and spin dynamics simulations have been performed to investigate the effect of vacancy defects on spin wave excitations in ferromagnetic iron.  Fourier transforms of space and time-displaced correlation functions yield the dynamic structure factor, providing characteristic frequencies and lifetimes of the spin wave modes.  Comparison of the system with a 5\% vacancy concentration with pure lattice data shows a decrease in frequency as well as a decrease in lifetime for all transverse spin wave excitations observed.  Additionally, a rugged spin wave line shape for low-$q$ spin waves indicates the presence of multiple localized excitations near defect sites resulting in reduced excitation lifetimes due to increased magnon-magnon scattering.  We observe further evidence of increased magnon-magnon scattering as the peaks in the longitudinal spin wave spectrum become less distinct.

\end{abstract}


\maketitle

\section{Introduction}
Magnetic properties of ferromagnetic metals at varying levels of impurity have previously been investigated using neutron scattering techniques\cite{lynn1975temperature, mook1973neutron, pickart1967spin, loong1984neutron, collins1969critical}. Notably, the introduction of substitutional defects into a magnetic system has been shown to distort the spectrum of the system's characteristic excitations\cite{svensson1969resonant}. This effect has been investigated both theoretically\cite{takeno1963spin, maradudin1965some, izyumov1966peculiarities} as well as experimentally for lattice vibrations and spin waves. These experiments have provided spin wave dispersion curves and stiffness constants. In particular, the spin wave stiffness parameter \textit{D} has shown sensitivity to variations in the concentration of magnetic defects\cite{antonini1969spin}. Additionally, both spin wave energies and lifetimes decrease as the impurity concentration grows.

Computational techniques have long been utilized to investigate collective excitations in systems spanning a broad range of practical interest. Molecular dynamics (MD) methods\cite{Rapaport} have been used to model vibrational behavior in metals, alloys, biological systems, etc. while spin dynamics (SD) simulations have investigated magnetic properties of classical lattice-based spin models\cite{landau1999spin}. SD simulations have replicated experimental findings in simple spin systems such as RbMnF\textsubscript{3}\cite{tsai2000spin} and have successfully predicted the existence of two-spin-wave modes\cite{bunker2000longitudinal}.  With the use of coordinate dependent exchange interactions, this method has also had success investigating systems with more complex magnetic properties such as propagating spin wave modes\cite{tao2005spin} and external magnetic field effects in bcc iron\cite{chui2014spin}.

Each of these methods numerically solves the classical equations of motion which describe the dynamical evolution of the model under consideration. Despite the individual capabilities of MD and SD simulation methods, the coupling of lattice and spin degrees of freedom is inherently neglected by both. In magnetic materials, atomic motion affects magnetic moments which depend non-trivially on the local atomic environment. Likewise, magnetic interactions have been shown to contribute to structural properties of these materials, including the BCC structure of iron\cite{Hasegawa1983, Herper1999}. Therefore, these degrees of freedom, atomic and magnetic, should be considered jointly in any model aiming to investigate the excitations of such a system.

The combined molecular dynamics - spin dynamics (MD-SD) method treats the spin subsystem using an extension of the Heisenberg model where the exchange interaction is a coordinate-dependent, pairwise function of inter-atomic distance\cite{Omelyan}. Inter-atomic interactions are handled with a non-magnetic many-body potential. The time integration of the coupled equations of motion is calculated using an algorithm based on the Suzuki-Trotter decomposition of the exponential time evolution operator. This algorithm is time reversible, efficient, and is known to conserve phase-space volume\cite{krech1998fast}. To handle systems of practical interest, such a model must utilize empirical many-body inter-atomic potentials and exchange interactions parameterized by experimental data and first-principles calculations\cite{Ma}. Using this compressible magnetic model parameterized for BCC iron, the interplay between spin waves (magnons) and lattice vibrations (phonons) has previously been investigated, showing spin wave dampening as well as magnon-phonon coupling in pure systems\cite{Perera}. 

This paper will extend the current MD-SD framework to investigate BCC iron with vacancies. We show the effect of these impurities on magnon dispersion curves as well as on individual spin wave excitation peaks. The spin waves are dampened and broadened in the presence of vacancy defects, in agreement with experimental neutron scattering data.



\section{Model}

\subsection{MD-SD Algorithm}
The MD-SD method extends the traditional MD approach through the addition of a third phase variable - the spin angular momentum $\mathbf{S}_i$ - to the position and velocity degrees of freedom. This spin variable is incorporated into the MD-SD Hamiltonian such that

\begin{equation}
\mathcal{H} = \sum_{i=1}^{N}{\frac{mv_{i}^{2}}{2}} + U(\{\bf{r}_{\it{i}}\}) - \sum_{\it{i} < \it{j}}{ \textnormal{J}_{\it{ij}}(\{\mathbf{r}_{\it{k}}\})}\bf{S}_{\it{i}} \cdot \bf{S}_{\it{j} } 
\end{equation}
\noindent
for a system of $N$ magnetic atoms of mass $m$ with positions $\{\mathbf{r}_{i}\}$, velocities $\{\mathbf{v}_{i}\}$, and atomic spins $\{\mathbf{S}_{i}\}$.  The first term in the MD-SD Hamiltonian represents the kinetic energy of the atoms, and $U(\{\mathbf{r}_{i}\})$ is a non-magnetic interatomic potential. The third term describes a Heisenberg-like exchange interaction which includes a coordinate-dependent exchange parameter $J_{ij}(\{\mathbf{r}_k\})$. The introduction of coordinate dependence into $J_{ij}(\{\mathbf{r}_{k}\})$ allows for the exchange interaction to depend on the locations of all atoms in the vicinity of atoms $i$ and $j$.  This framework is general and may be utilized for any magnetic material given proper parameterization of the interatomic potential and exchange parameter. To investigate BCC iron, we employ the embedded atom interatomic potential $U(\{r_{i}\})$ developed and parameterized for iron by Dudarev and Derlet \cite{Dudarev}. The exchange interaction used in this work is a pairwise function $J(r_{ij})$ parameterized by first principles calculations for iron \cite{Ma}. 

This MD-SD Hamiltonian has true dynamics according to the classical equations of motion

\begin{subequations}\label{eom}
\begin{align}
&\frac{d\bf{r_{\it{i}}}}{dt} = \bf{v}_{\it{i}}\\[5pt]
&\frac{d\bf{v_{\it{i}}}}{dt} = \frac{\bf{f_{\it{i}}}}{m} \\[5pt]
&\frac{d\bf{S_{\it{i}}}}{dt} = \frac{1}{\hbar}\bf{H}_{\it{i}}^{\it{eff}} \times \bf{S}_{\it{i}} 
\end{align}
\end{subequations}
\noindent
where $\mathbf{f}_i$ and $\mathbf{H}^{eff}_{i}$ represent the interatomic force and effective field acting on the $i^{th}$ atom, respectively. 

In order to investigate the collective excitations of the simulated system, space-displaced, time-displaced correlation functions are computed. By performing Fourier transformations of these correlation functions, we obtain information about the frequency and lifetimes of these excitations.  During a simulation run, the spatial Fourier transform of the space-displaced, time-displaced spin-spin correlation function, or the intermediate scattering function, is calculated on the fly

\begin{equation}
F_{ss}^{k}(\mathbf{q},t)=\frac{1}{N}\langle \rho_{s}^{k}(\mathbf{q},t)\rho_{s}^{k}(-\mathbf{q},0) \rangle,
\end{equation}
\noindent
where $k$ represents the real-space directions $\{x, y, z\}$ and $\rho_{s}(\mathbf{q}, t)$ represents the microscopic spin density, defined as
\begin{equation}
 \rho_{s}(\mathbf{q}, t) = \sum_{i}\mathbf{S}_{i}(t)e^{-i\mathbf{q} \cdot \mathbf{r}_{i}(t)}. 
\end{equation}
During a microcanonical simulation of a ferromagnetic material, the total magnetization vector is a constant of motion. This property allows for the differentiation of spin excitations which propagate parallel and perpendicular to the magnetic symmetry axis through the choice of a Cartesian coordinate system such that the $z$-axis is parallel to the magnetization vector. The components of $F_{ss}(\mathbf{q}, t)$ are then regrouped into a longitudinal component

\begin{equation}
F_{ss}^{L}(\mathbf{q}, t) = F_{ss}^{z}(\mathbf{q}, t)
\end{equation}
\noindent
and a transverse component
\begin{equation}
F_{ss}^{T}(\mathbf{q}, t) = \frac{1}{2} [ F_{ss}^{x}(\mathbf{q}, t) + F_{ss}^{y}(\mathbf{q}, t) ].
\end{equation}
\noindent
A temporal Fourier transform of the intermediate scattering function yields the spin-spin dynamic structure factor
\begin{equation}
S_{ss}^{L,T}(\mathbf{q}, \omega) = \frac{1}{2\pi}\int_{-\infty}^{+\infty}F_{ss}^{L,T}(\mathbf{q}, t)e^{-i\omega t}dt
\end{equation}
\noindent
for momentum transfer $\mathbf{q}$ and frequency $\omega$. The dynamic structure factor obtained from MD-SD simulations may be compared to the measurements made by inelastic neutron scattering experiments\cite{lovesey1984theory}.

\subsection{Simulation Details}

Time integration of the equations of motion shown in Eq. (\ref{eom}) is performed using a second-order Suzuki-Trotter decomposition of the time evolution operator \cite{Omelyan, tsai2005molecular}. We equilibrate the position and spin subsystems of multiple initial configurations to the desired temperature $T$ using the Metropolis Monte Carlo method. Initial velocities are then drawn from the Maxwell-Boltzmann distribution at this temperature. A short microcanonical MD-SD run is performed in order to equilibrate the three sub-systems. Once properly equilibrated, these configurations are independently time-evolved using the microcanonical MD-SD integration scheme. For this time integration, a time-step of $\delta t = 1$ fs was found to adequately conserve energy and magnetization. Simulation data shown in this paper are from MD-SD runs of 1 ns ($10^{6}$ time-steps) in duration. During the time integration, the intermediate scattering function is recorded on the fly for each independent configuration. Once all runs have completed we average these data, yielding an estimate of the canonical ensemble average at temperature $T$. 


\begin{figure}
\includegraphics[width=0.5\textwidth]{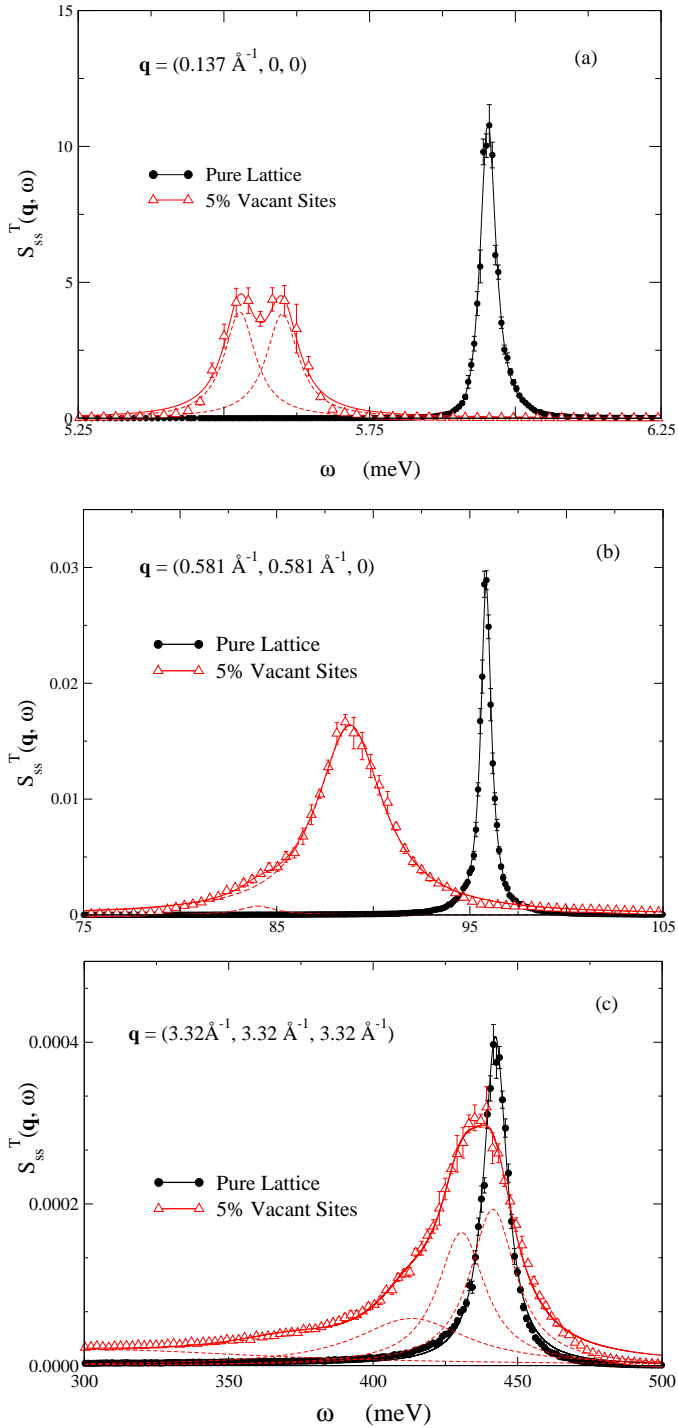}
\caption{\label{Spin Wave Peaks} Transverse spin-spin dynamic structure factor obtained from MD-SD simulations for L = 16 at $T$ = 300 K for both the pure lattice and the system with 5\% of lattice sites left vacant. (a) $\mathbf{q} = (0.137 \angstrom^{-1}, 0, 0)$, (b) $\mathbf{q} = (0.581 \angstrom^{-1}, 0.581 \angstrom^{-1}, 0)$, (c) $\mathbf{q} = (3.32 \angstrom^{-1}, 3.32 \angstrom^{-1}, 3.32 \angstrom^{-1})$. Symbols represent simulation data while solid lines show multi-peak Lorentzian curve-fitting.  Dashed lines indicate the constituent peaks which make up the solid curve for the systems with 5\% vacancy.}
\end{figure}

\begin{figure}
\includegraphics[width=0.5\textwidth]{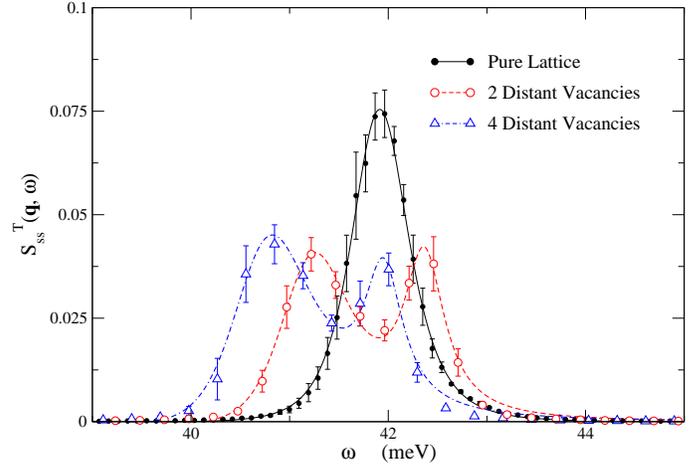}
\caption{\label{Spin Wave Peaks} Transverse spin-spin dynamic structure factor obtained from MD-SD simulations for L = 6 at $T$ = 300 K for $\mathbf{q} = (0.365 \angstrom^{-1}, 0, 0)$. Curves with symbols represent configurations which have 2 or 4 atoms removed from the system. }
\end{figure}

In order to include vacancy type defects, we leave randomly selected sites devoid of an atom in each initial configuration. For the data presented here, simulations have been performed for a body-centered cubic system of edge length $L$ = 16 (8192 sites) with periodic boundary conditions at $T$ = 300 K. For a 5\% vacancy concentration, 7783 sites contain an atom while the remaining 409 sites remain vacant. Each independent configuration included in the overall simulation set has a unique vacancy distribution. 

Using the canonical ensemble average estimate for the spin-spin intermediate scattering function obtained from these simulations, we calculate the dynamic structure factor.  Spin wave excitations are observed through the transverse spin-spin dynamic structure factor $S^{T}_{ss}(\mathbf{q}, \omega)$ as peaks of Lorentzian form,
\begin{equation}\label{lorentzian}
S^{T}_{ss}(\mathbf{q}, \omega) = \frac{I_{0}\Gamma^{2}}{(\omega - \omega_{0})^{2}+\Gamma^{2}} 
\end{equation}
\noindent
where $\omega_{0}$ is the characteristic frequency of the excitation, $I_{0}$ is the amplitude, and $\Gamma$ is the half width at half maximum.  $\Gamma$ is inversely proportional to the lifetime of the associated excitation.  Fitting simulational data to Eq. 8 allows us to observe the impact of vacancy defects on spin wave dispersion curves, as well as on individual spin wave line shapes, excitation energies, and lifetimes.

\section{Results}

\subsection{Transverse Spin Waves}

\subsubsection{Spin Wave Energies}

\begin{figure}
\includegraphics[width=0.5\textwidth]{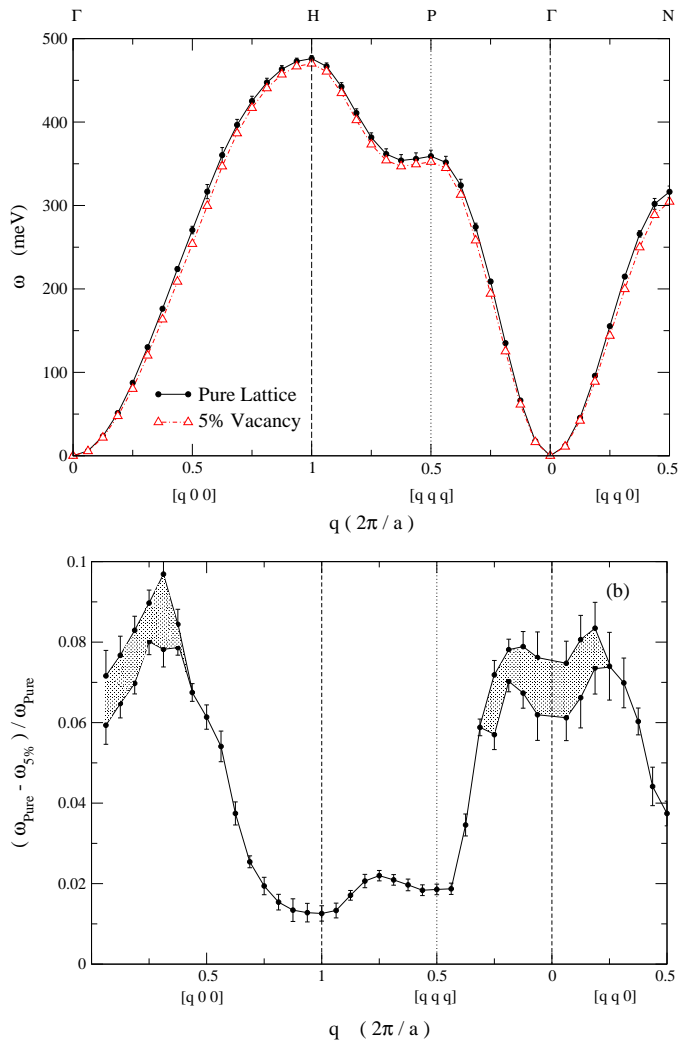}
\caption{\label{Magnon Shift} (a) Spin wave dispersion curve at $T=300$ K obtained from MD-SD simulations with $L=16$.  Results are shown for the pure lattice as well as the system with 5\% vacancy concentration.  (b) The fractional shift in transverse spin wave frequencies due to a 5\% concentration of vacancy defects. Results obtained from MD-SD simulations of systems of size $L = 16$ at $T=300$ K.}
\end{figure}

In order to demonstrate our approach, we performed calculations with a 5\% vacancy concentration, a figure consistent with the nonmagnetic defect concentration found in experimental studies.  Figure 1 shows the effect of this 5\% vacancy concentration on three different transverse spin waves in a system of size $L=16$.  The vacancy defects cause a dampening of the peaks in $S_{ss}^{T}(\mathbf{q}, \omega)$ as well as a shift to lower frequency.  These effects on spin wave peaks are observed for all wave vectors, though we only show a few selected line shapes here.  Visible in Fig. 1(a), the low-$q$ peak region of the impure system displays a more rugged spectral structure than the pure peaks or any of the higher $q$ peaks from impure systems.  While curve fitting these rough peaks to the form of Eq. (\ref{lorentzian}) yields estimates of peak locations and half-widths, applying a fit utilizing the sum of multiple Lorentzian forms more accurately reproduces the asymmetries and additional structure in these peaks.  
Included in Fig. 1 are fits to two-peak (Figs. 1(a), 1(b)) and four-peak (Fig. 1(c)) Lorentzian functions, describing the distorted signals obtained from the impure system.  Multiple peaks are also visible in the small-$q$ region in the $(q, q, 0)$ and $(q, q, q)$ symmetry directions, though we only show the $(q, 0, 0)$ peak in Fig. 1.  The distortion of the these spin wave line shapes is due to the presence of localized spin wave modes near the defect sites.  While these multi-peak Lorentzian functions provide reasonable fits to our data, the possibility remains of additional structure within the limits of our resolution.

In order to investigate the downward shift of spin wave frequencies seen in Fig. 1, we simulated a small system ($L = 6$) from which atoms are removed one by one.  The decrease in spin wave frequency due to this removal of atoms is observed in Figure 2.  The atoms removed from the configurations in Fig. 2 are chosen such that the sites are separated by a distance greater than the cutoff distance of the interatomic potential.  Multiple vacancy configurations were generated, though only one configuration is shown for each data set in Fig. 2.  As the number of vacancy sites grows, the frequency of the characteristic spin wave spectrum decreases.

The full spin wave dispersion curve is shown in Figure 3(a), constructed using the peak locations obtained through Lorentzian curve fitting.  To observe the effect of vacancy defects on spin wave excitations more carefully, we compare the characteristic spin wave frequencies obtained from MD-SD simulations for the pure system with those from systems with a 5\% vacancy concentration.  We calculate the fractional shift in spin wave frequency, $(\omega_{\textrm{\scriptsize{ Pure}}} - \omega_{\textrm{\scriptsize{ 5\%}}} / \omega_{\textrm{\scriptsize{ Pure}}})$, shown in Fig. 3(b) directly below the dispersion curve.  All observed spin wave excitations display a shift to lower frequency, with this effect being most significant near the zone center.  The shaded areas in Fig. 3(b) indicate the regions in the small-\textit{q} $S_{ss}^{T}(\mathbf{q}, \omega)$ spectra where we observe multiple peaks.  While multi-peak Lorentzian forms fit the high-\textit{q} signals more accurately than the single peak form, identification of the less intense constituent peaks is unreliable.  Therefore we include only the most intense peak signal from our curve-fitting procedure for high-\textit{q} excitations in Fig. 3(b).

\begin{figure}
\includegraphics[width=0.45\textwidth]{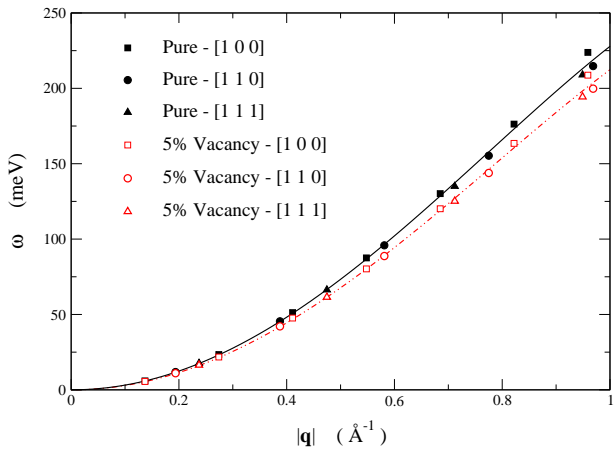}
\caption{\label{Low-q Magnon dispersion} Comparison of low $|\mathbf{q}|$ magnon dispersion curves obtained from MD-SD simulation (L=16) for the pure system and that with 5\% vacancy defects. Lines in the figure represent a two-parameter curve fit to Eq. 9 with $D$ shown in Table I.}
\end{figure}

In ferromagnetic materials below the critical temperature $T_{c}$, spin waves with small-$q$ are isotropic and are expected to approximate a quadratic dispersion relation\cite{shirane1968spin} of the form 
\begin{equation}\label{dispersion}
\hbar\omega = D|\mathbf{q}|^2(1-\beta |\mathbf{q}|^2)
\end{equation}
\noindent
where $D$ represents the magnetic stiffness constant.  The dispersion curve for low-$q$ is shown in Figure 4 for both the pure system and that with 5\% vacancy concentration.  The dispersion parameters $D$ and $\beta$ in Eq. (\ref{dispersion}) are quantities accessible through neutron scattering using diffraction methods (DM), triple axis spectrometry (TAS), chopper spectrometry (CS), or other methods.  While the dispersion relation shown in Eq. (\ref{dispersion}) provides an accurate fit to both experimental and computational data, the behavior of the parameter $\beta$ as observed in experiment is inconsistent.  However previous investigation of BCC iron under varying levels of impurity has shown systematic behavior in the stiffness parameter $D$.  The conditions most notable to this study are the substitutional inclusion of Si, a nonmagnetic atom, into the crystal.  Due to the lack of a magnetic moment and lower mass in comparison to Fe, the Si atom mimics a magnetic vacancy.  The lines in Fig. 4 represent curve fits to Eq. (\ref{dispersion}), and the values of the stiffness fit parameter $D$ are shown along with previous experimental data in Table \ref{stiffness_table}.  While our results overestimate the value of the stiffness parameter, they capture the trend of a decrease in $D$ caused by nonmagnetic defects observed in experimental data.

\subsubsection{Spin Wave Lifetimes}
\begin{table}[]
\renewcommand{\arraystretch}{1.5}
\centering
\caption{Experimental and computational comparison of spin wave dispersion parameters in impure Fe systems at room temperature. }
\label{stiffness_table}
\begin{tabular*}{\columnwidth}{@{\extracolsep{3mm} }ccccc}
 & D (meV \angstrom \textsuperscript{2}) & Method & Ref. \\
\Xhline{2.5\arrayrulewidth}
Fe              & 307 $\pm$ 15  &  CS         & \citet{loong1984neutron}\\ \hline
Fe               & 281 $\pm$ 10 & \multirow{2}{*}{ TAS}       & \multirow{2}{*}{ \citet{shirane1968spin}}\\
Fe (4\% Si)            & 270 $\pm$ 10  & \\ \hline 
Fe               & 315.1 $\pm$ 0.1      & \multirow{2}{*}{ MD-SD}   & \multirow{2}{*}{ This work}\\
Fe (5\% Vac.)       & 290.2 $\pm$ 0.2  \\\hline
Fe (7\% Si)       & 273 $\pm$ 3     & \multirow{2}{*}{ DM}      &\multirow{2}{*}{ \citet{antonini1969spin}}\\
Fe (15\% Si)           & 233 $\pm$ 11   

\end{tabular*}
\end{table}

\begin{figure}
\includegraphics[width=0.5\textwidth]{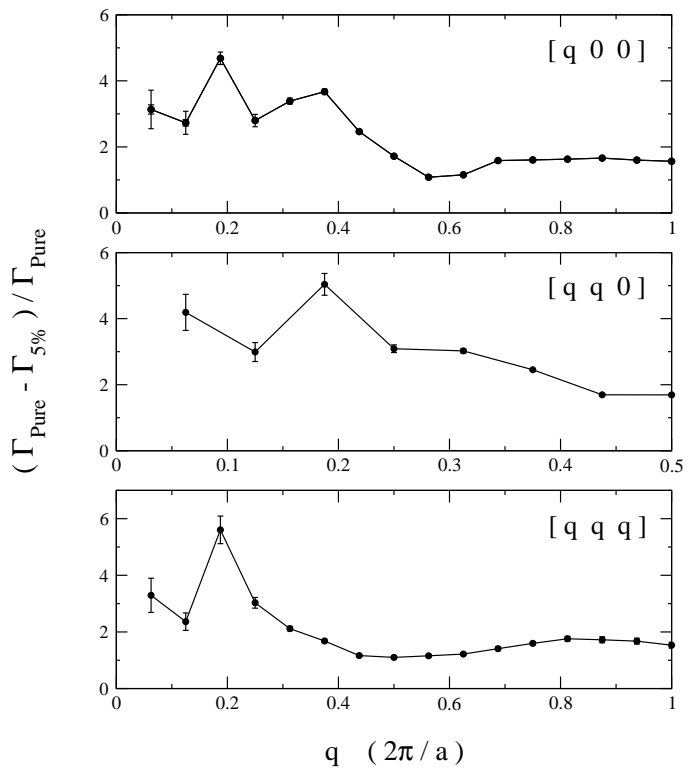}
\caption{\label{Magnon HWHM 3 Directions} Fractional shift in half width at half maximum ($\Gamma$) of transverse spin waves at $T$ = 300 K obtained from MD-SD simulations for $L=16$ in the [1 0 0], [1 1 0], and [1 1 1] symmetry directions.  }
\end{figure}

\begin{figure}
\includegraphics[width=0.45\textwidth]{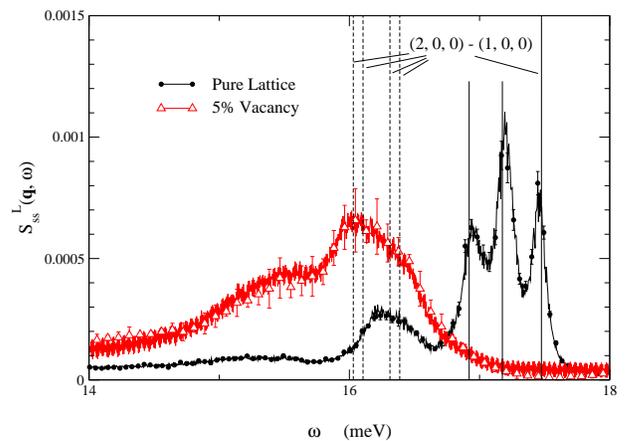}
\caption{\label{Longitudinal} Longitudinal spin-spin dynamic structure factor obtained from MD-SD simulations for $\mathbf{q} = (0.137 \angstrom^{-1}, 0, 0)$ at $T$ = 300 K with $L = 16$ for both the pure system and that with 5\% of sites left vacant. Solid lines represent the predicted peak locations in the pure system, and dashed lines represent the predicted locations of the $(2, 0, 0) - (1, 0, 0)$ peaks in the system with vacancies.  For clarity, only the $(2, 0, 0) - (1, 0, 0)$ excitations are labeled.}
\end{figure}

Spin wave excitation lifetimes are inversely proportional to the half width at half maximum of the characteristic peak in $S^{T}_{ss}(\mathbf{q}, \omega)$ which may be obtained through curve fitting to Eq. (\ref{lorentzian}).  The effect of vacancy defects on spin wave lifetime is presented in Figure 5, which shows the fractional change in the half width at half maximum due to impurities.  Fig. 5 shows significant broadening of spin wave peaks for all $q$ values, indicating a decrease in lifetime for all spin wave excitations.  However the effect is most significant for low $q$ excitations, and the broadening decreases as $q$ grows larger in all symmetry directions.  This decreased lifetime at low $q$ is due to increased magnon-magnon scattering caused by the existence of additional spin wave modes, evidenced by the rugged structure of low $q$ peaks in $S^{T}_{ss}(\mathbf{q}, \omega)$ as in Fig. 1 (a).

\subsection{Longitudinal Spin Waves}
For classical Heisenberg spin models, Bunker \textit{et al.} have previously shown that the excitation peaks in the longitudinal component of $S_{ss}(\mathbf{q}, \omega)$ represent creation and/or annihilation processes resulting from the interaction of multiple transverse spin waves\cite{bunker2000longitudinal}.  Previous MD-SD simulation of pure BCC iron measured the longitudinal component of $S_{ss}(\mathbf{q}, \omega)$, and the frequencies of these two-spin wave peaks were identified using the difference of transverse spin wave frequencies\cite{Perera}:

\begin{equation}
    \omega_{ij}^{-}(\mathbf{q}_{i}\pm \mathbf{q}_{j}) = \omega(\mathbf{q}_{i}) - \omega(\mathbf{q}_{j})
\end{equation}

\noindent where $\mathbf{q}_{i}$ and $\mathbf{q}_{j}$ are the wave-vectors of the constituent spin waves, and $\omega$ is the characteristic frequency of each.

Figure 6 shows a portion of the $S_{ss}^{L}(\mathbf{q}, \omega)$ spectrum for $\mathbf{q} = \frac{2\pi}{La}(1, 0, 0)$ the $L = 16$ system.  Since the large set of available wave vectors $\{\mathbf{q}_{i}\}$ in an $L = 16$ system leads to many two-spin-wave interaction processes, only a small section of $S_{ss}^{L}(\mathbf{q}, \omega)$ is presented here.  Individual peaks in the pure lattice spectrum are well defined and we have identified the two transverse spin waves that combine to produce each peak in the longitudinal component.  An example of the identification of these peaks is shown by the solid vertical line in Fig. 6 which represents the difference in frequencies of the $\frac{2\pi}{La}(1, 0, 0)$ and $\frac{2\pi}{La}(2, 0, 0)$ transverse spin waves, calculated using Eq. 10.  While we have identified the other peaks in the pure lattice data of Fig. 6, only one is shown in the figure for clarity.

Fig. 6 also shows the effect of 5\% vacancy defects on $S_{ss}^{L}(\mathbf{q}, \omega)$.  Qualitatively, the longitudinal spectrum is less well-defined for the impure system, as the individual excitation peaks seen in the pure system shift to lower frequency and merge into a broad distribution.  As shown previously in Fig. 1(a), small-$q$ spin waves such as those that contribute to the solid line in Fig. 6 display multiple transverse excitation peaks.  Therefore to identify the location of longitudinal excitations in the impure system, we must consider each combination of the constituent transverse spin waves.  Both of the $\frac{2\pi}{La}(1, 0, 0)$ and $\frac{2\pi}{La}(2, 0, 0)$ transverse spin wave spectra show dual-peak structure, therefore Eq. 10 has at least four unique combinations of spin wave interactions.  Each of these resultant frequencies are shown as dashed vertical lines in Fig. 6, indicating that the loss of individual peak resolution in this region of $S_{ss}^{L}(\mathbf{q}, \omega)$ is due to the increase in multiple spin wave annihilation processes.

\section{Conclusions}
The effect of vacancy defects on magnetic excitations in BCC iron were investigated using combined molecular and spin dynamics simulations (MD-SD) at room temperature.  We calculated the intermediate scattering function on-the-fly during our simulation runs, and used this data to obtain the dynamic structure factor $S_{ss}(\mathbf{q},\omega)$ which contains information regarding spin wave energies and lifetimes.  We obtained spin wave energies and half widths at half maximum by performing curve-fitting of the characteristic spin wave peaks to a Lorentzian function.  For all observed spin waves, the introduction of vacancy defects was shown to decrease the energy of the excitation with the effect being most significant near the zone center.  The $S_{ss}(\mathbf{q},\omega)$ line shape in low-$q$ spin waves also showed rugged structure, indicating the propagation of multiple excitations with different characteristic energies.  

The presence of multiple spin waves modes increases magnon-magnon scattering, which is evident by the decrease in observed spin wave lifetime for low-$q$ excitations.  The low $q$ region of the magnon dispersion curve has been shown experimentally to obey a quadratic dispersion relation of the form of Eq. (\ref{dispersion}).  From this quadratic form, the magnetic stiffness constant has been measured experimentally for BCC iron under varying conditions of magnetic impurity.  Our findings using MD-SD simulations are consistent with the observed decrease in the stiffness constant with the introduction of defects, as shown in Fig. 4 and Table I.  We investigated longitudinal spin wave modes which represent two spin wave interaction processes.  The introduction of defects into the system resulted in the loss of clearly defined peaks which are observed in the pure system.  We have calculated the resultant frequency of interacting transverse spin waves, showing that this effect is due to an increase in available spin wave scattering processes.  Further quantitative studies of magnetic materials containing impurity atoms (e.g. Fe\textsubscript{$1-x$}Si\textsubscript{$x$}) will require reliable interatomic potentials for those alloys.

\vspace{6mm}

\section*{Acknowledgments}
Part of this work (M.E.) was supported by the U.S. Department of Energy, Office of Science, Basic Energy Sciences, Material Sciences and Engineering Division.  This work was sponsored in part by the “Center for Defect Physics”, an Energy Frontier Research Center of the Office of Basic Energy Sciences (BES), U.S. Department of Energy (DOE).  This study was supported in part by resources and technical expertise from the Georgia Advanced Computing Resource Center, a partnership between the University of Georgia's Office of the Vice President for Research and Office of the Vice President for Information Technology.  This research also used resources of the Oak Ridge Leadership Computing Facility at ORNL, which is supported by the Office of Science of the U.S. Department of Energy under Contract No. DE-AC05-00OR22725.

\bibliography{RandomVacancies.bib}

\end{document}